# Measuring the entanglement of analogue Hawking radiation by the density-density correlation function

Jeff Steinhauer

*Department of Physics, Technion—Israel Institute of Technology, Technion City, Haifa 32000, Israel*

We theoretically study the entanglement of Hawking radiation pairs emitted by an analogue black hole. We find that this entanglement can be measured by the experimentally accessible density-density correlation function, vastly simplifying the measurement. We find that while the Hawking radiation exiting the black hole might be Planck-distributed, the correlations between the Hawking radiation and the partner particles has a distribution which is weaker but broader than Planckian. Thus, the high energy tail of the distribution of Hawking radiation should be entangled, whereas the low energy part should not be. This confirms a previous numerical study. The full Peres-Horodecki criterion is considered, as well as a simpler criterion in the stationary, homogeneous case. Our method applies to systems which are sufficiently cold that the thermal phonons can be neglected.

The fate of information as a black hole evaporates is a puzzle with a variety of proposed solutions [1-3]. An important element of this puzzle is the entanglement between the infalling and outgoing members of a Hawking pair. Verifying whether this entanglement actually exists seems very difficult since one should observe both sides of the black hole horizon. The extreme weakness of the Hawking radiation adds further difficulty. It thus falls upon the field of analog gravity to shed light on this issue. It was proposed that an analog black hole could be used to observe Hawking radiation [4]. This observation was achieved recently [5]. It was further proposed that the entanglement of the analog Hawking radiation could be measured through a series of non-destructive measurements in an optical cavity [6]. Here, we present a significant simplification, by showing that the entanglement could be observed through the density-density correlation function. The latter is readily observable through destructive *in situ* imaging [5].

Hawking radiation in Bose-Einstein condensates has been studied extensively theoretically [6-18]. The density-density correlation function was presented as an important measurement tool [10]. Recently, theoretical attention has turned toward the entanglement of the analog Hawking radiation [6,16-18].

Experimentally, self-amplifying Hawking radiation in an analogue black hole was recently observed [5]. Previously, surface waves in water were used to observe mode conversion by an analogue white hole horizon [19]. Studies are underway in other systems such as nonlinear optical fibers [20-22] and exciton-polariton condensates [23].

Here, we show that the entanglement of the Hawking radiation can be measured from the density-density correlation function. We rely on the assumption that the population of the thermal phonons is negligible. The entanglement can then be observed with only one observation for each member of the ensemble, allowing for the use of standard *in situ* imaging. The technique is a generalization of our *in situ* Fourier transform technique for measuring phonon populations [24], which allowed us to prove that the initial state of our analog black hole was quantum in nature [5]. Our technique is applicable to an analog black hole with homogeneous upstream and downstream regions. The upstream (downstream) region has subsonic (supersonic) flow in the $-x$ direction. The relevant Bogoliubov excitations have positive wavenumber $k$, and travel against the flow in the local rest frame of the fluid.

The entanglement can be determined by the Peres-Horodecki criterion discussed at the end of this work. For clarity however, we will usually focus on the following more simple measure of the nonseparability of the state [16],

$$\Delta \equiv \langle \hat{b}^{u\,\dagger}_{k_{HR}} \hat{b}^{u}_{k_{HR}} \rangle \langle \hat{b}^{d\,\dagger}_{k_P} \hat{b}^{d}_{k_P} \rangle - \left| \langle \hat{b}^{u}_{k_{HR}} \hat{b}^{d}_{k_P} \rangle \right|^2, \qquad (1)$$

where $\hat{b}^{u}_{k_{HR}}$ is the annihilation operator for a Bogoliubov excitation with wavenumber $k_{HR}$, localized in the subsonic upstream region (outside the black hole). In other words, $u, d, HR$ and $P$ stand for "upstream", "downstream", "Hawking radiation", and "partner", respectively. If $\Delta$ is negative, then the correlations between the Hawking and partner particles are strong enough to indicate that they are entangled.

We propose to measure $\Delta$ in (1) by *in situ* imaging, which destructively measures the densities in the upstream and downstream regions. Firstly, we will write $\Delta$ in terms of the Fourier transform of the density operator [25]

$$\rho_k = \sum_p \hat{a}^\dagger_{p+k} \hat{a}_p \quad (2)$$

where $\hat{a}_p$ is the annihilation operator for a single atom with momentum $\hbar p$. In the Bogoliubov approximation, this can be written [25]

$$\rho_k = \sqrt{N}(u_k + v_k)(\hat{b}^\dagger_k + \hat{b}_{-k})$$

where $N$ is the total number of atoms and $u_k$ and $v_k$ are the Bogoliubov amplitudes. We can treat the upstream and downstream regions separately and write

$$\rho^u_k = \sqrt{N^u}(u_{k_{HR}} + v_{k_{HR}})(\hat{b}^{u\,\dagger}_{k_{HR}} + \hat{b}^u_{-k_{HR}})$$

where $N^u$ is the total number of atoms in the upstream region, and $u_{k_{HR}}$ and $v_{k_{HR}}$ are the Bogoliubov coefficients computed with the homogeneous upstream density, which is not necessarily the same as the homogeneous downstream density. Similarly,

$$\rho^d_k = \sqrt{N^d}(u_{k_P} + v_{k_P})(\hat{b}^{u\,\dagger}_{k_P} + \hat{b}^u_{-k_P}).$$

We can now define a generalized version of the usual static structure factor given by $S(k) = N^{-1}\langle \rho_k \rho_{-k}\rangle$ in a homogeneous system.

$$\langle \rho^i_{k_i} \rho^j_{k_j}\rangle = \sqrt{N^i N^j}(u_{k_i} + v_{k_i})(u_{k_j} + v_{k_j})\left[\langle \hat{b}^{i\,\dagger}_{k_i} \hat{b}^{j\,\dagger}_{k_j}\rangle + \langle \hat{b}^{i\,\dagger}_{k_i} \hat{b}^j_{-k_j}\rangle + \langle \hat{b}^i_{-k_i} \hat{b}^j_{k_j}\rangle + \langle \hat{b}^i_{-k_i} \hat{b}^j_{-k_j}\rangle + \delta_{ij}\delta_{-k_i k_j}\right] \quad (3)$$

where $i$ and $j$ can each be either $u$ or $d$. The last term comes from the bosonic commutation relation.

We would now like to construct $\Delta$ given in (1). Firstly, consider the case $i = u$, $j = d$, $k_i = -k_{HR}$, and $k_j = -k_P$. Eq. 3 then becomes

$$\langle \rho^u_{-k_{HR}} \rho^d_{-k_P}\rangle = \sqrt{N^u N^d}(u_{k_{HR}} + v_{k_{HR}})(u_{k_P} + v_{k_P})\left[\langle \hat{b}^{u\,\dagger}_{-k_{HR}} \hat{b}^{d\,\dagger}_{-k_P}\rangle + \langle \hat{b}^{u\,\dagger}_{-k_{HR}} \hat{b}^d_{k_P}\rangle + \langle \hat{b}^{u\,\dagger}_{k_{HR}} \hat{b}^d_{-k_P}\rangle + \langle \hat{b}^u_{k_{HR}} \hat{b}^d_{k_P}\rangle\right]$$

We now assume that the number of excitations traveling with the flow (with negative $k$) is negligible. Such excitations would result from the finite temperature of the condensate. We are thus assuming that this temperature is sufficiently low. Thus, we can set any term with a negative $k$-value to zero. We thus obtain

$$\langle \rho^u_{-k_{HR}} \rho^d_{-k_P}\rangle = \sqrt{N^u N^d}(u_{k_{HR}} + v_{k_{HR}})(u_{k_P} + v_{k_P})\langle \hat{b}^u_{k_{HR}} \hat{b}^d_{k_P}\rangle. \quad (4)$$

This is proportional to the first correlation function in (1). We can similarly obtain the other correlation functions by considering two cases with $i = j$ in (3). In the upstream region,

$$\langle \rho^u_{k_{HR}} \rho^u_{-k_{HR}} \rangle = N^u (u_{k_{HR}} + v_{k_{HR}})^2 [\langle \hat{b}^{u\dagger}_{k_{HR}} \hat{b}^u_{k_{HR}} \rangle + 1]. \tag{5}$$

In the downstream region,

$$\langle \rho^d_{k_P} \rho^d_{-k_P} \rangle = N^d (u_{k_P} + v_{k_P})^2 [\langle \hat{b}^{d\dagger}_{k_P} \hat{b}^d_{k_P} \rangle + 1]. \tag{6}$$

Note that (5) and (6) are the usual static structure factor, and (4) is a generalized static structure factor. By inserting (4), (5), and (6) in (1),

$$\Delta = [F_2 \langle \rho^u_{k_{HR}} \rho^u_{-k_{HR}} \rangle - 1][F_3 \langle \rho^d_{k_P} \rho^d_{-k_P} \rangle - 1] - F_1 |\langle \rho^u_{-k_{HR}} \rho^d_{-k_P} \rangle|^2 \tag{7}$$

where the $F$ factors are smoothly varying functions of $k$ which depend on the parameters of the upstream and downstream regions,

$$F_1 \equiv \left[ \sqrt{N^u N^d} (u_{k_{HR}} + v_{k_{HR}})(u_{k_P} + v_{k_P}) \right]^{-2}$$

$$F_2 \equiv \left[ N^u (u_{k_{HR}} + v_{k_{HR}})^2 \right]^{-1}$$

$$F_3 \equiv \left[ N^d (u_{k_P} + v_{k_P})^2 \right]^{-1}$$

where $(u_k + v_k)^2$ is derivable from the product of $k$ and the healing length $\xi = \hbar/mc$, where $c$ is the speed of sound in the relevant region.

We see that (7) gives us the desired entanglement criterion in terms of the $\rho_k$ operators. We used the 2nd quantized form (2) of $\rho_k$ to derive this expression. Now, to connect with the experimental images, we will use the expression for $\rho_k$ explicitly as a Fourier transform,

$$\rho^i_{k_i} = \int dx e^{-ik_i x} n^i(x)$$

where $n^u(x)$ and $n^d(x)$ are the densities in the upstream and downstream regions, respectively. Thus,

$$\langle \rho^i_{k_i} \rho^j_{k_j} \rangle = \int dx_1 dx_2 e^{-ik_i x_1} e^{-ik_j x_2} \langle n^i(x_1) n^j(x_2) \rangle$$

We thus see that the $\langle \rho^i_{k_i} \rho^j_{k_j} \rangle$ can be measured by computing the Fourier transform of the experimental density-density correlation function. Choosing the parameters as relevant for (7),

$$\langle \rho^u_{-k_{HR}} \rho^d_{-k_P} \rangle = \int dx_1 dx_2 e^{ik_{HR}x_1} e^{ik_P x_2} \langle n^u(x_1) n^d(x_2) \rangle \tag{8}$$

$$\langle \rho^u_{k_{HR}} \rho^u_{-k_{HR}} \rangle = \int dx_1 dx_2 e^{-ik_{HR}x_1} e^{ik_{HR}x_2} \langle n^u(x_1) n^u(x_2) \rangle \tag{9}$$

$$\langle \rho^d_{k_P} \rho^d_{-k_P} \rangle = \int dx_1 dx_2 e^{-ik_P x_1} e^{ik_P x_2} \langle n^d(x_1) n^d(x_2) \rangle \tag{10}$$

We see that the $\langle \rho^i_{k_i} \rho^j_{k_j} \rangle$ needed for computing $\Delta$ are given by the Fourier transforms of various areas of the position-space correlation function, as shown in Fig. 1. The horizon is at the origin, and the quadrants $(i,j) = (u,u)$, etc. are labeled.

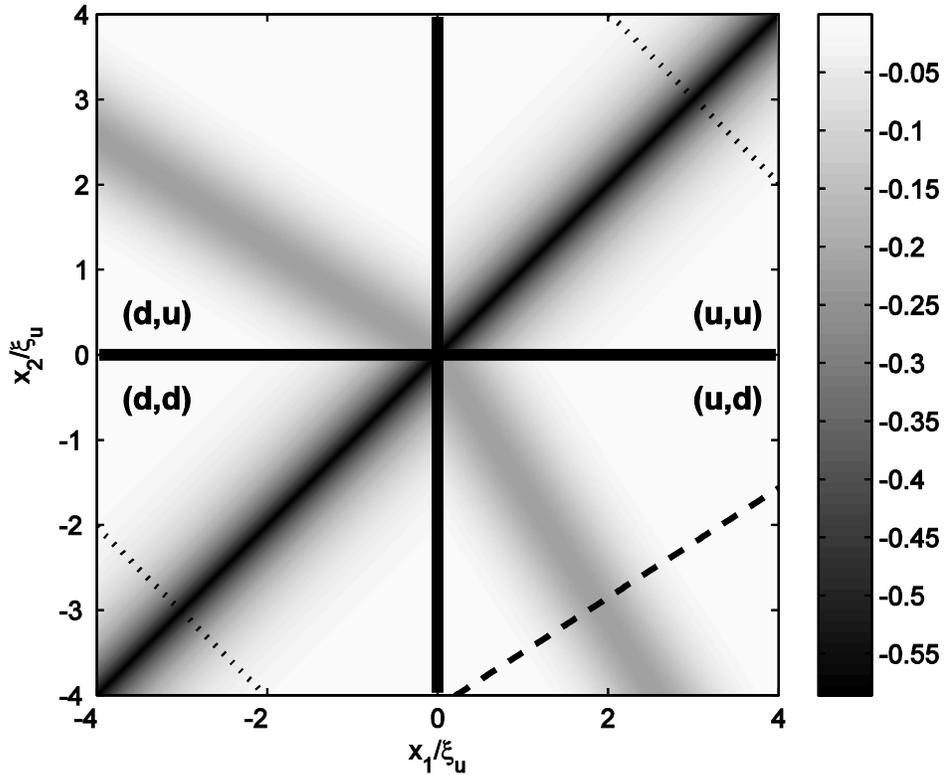

Fig. 1. The quadrants in position space. The (d, u) and (u, d) quadrants show the analytic result of Ref. 10. The profile along the dotted line in the (u, u) or (d, d) quadrants is the Fourier transform of the usual static structure factor. Comparing the Fourier transforms along the dotted and dashed lines gives a measure of the entanglement. For simplicity, the delta function along the $x_1 = x_2$ diagonal is not shown. This delta function merely adds a constant when the Fourier transform along the dotted line is computed. $x_1$ and $x_2$ are normalized by the upstream healing length $\xi_u$.

By (8), the Fourier transform of the $(u,d)$ quadrant should be computed in order to obtain $\langle \rho^u_{-k_{HR}} \rho^d_{-k_P} \rangle$, which then gives $\langle \hat{b}^u_{k_{HR}} \hat{b}^d_{k_P} \rangle$ by (4). This Fourier transform is shown in Fig. 2. The linear feature seen gives the corresponding values of the wavenumbers of Hawking-partner pairs. Equivalently, one can compute the 1D Fourier transform along the dashed line of Fig. 1, as indicated by the dashed curve in Fig. 3.

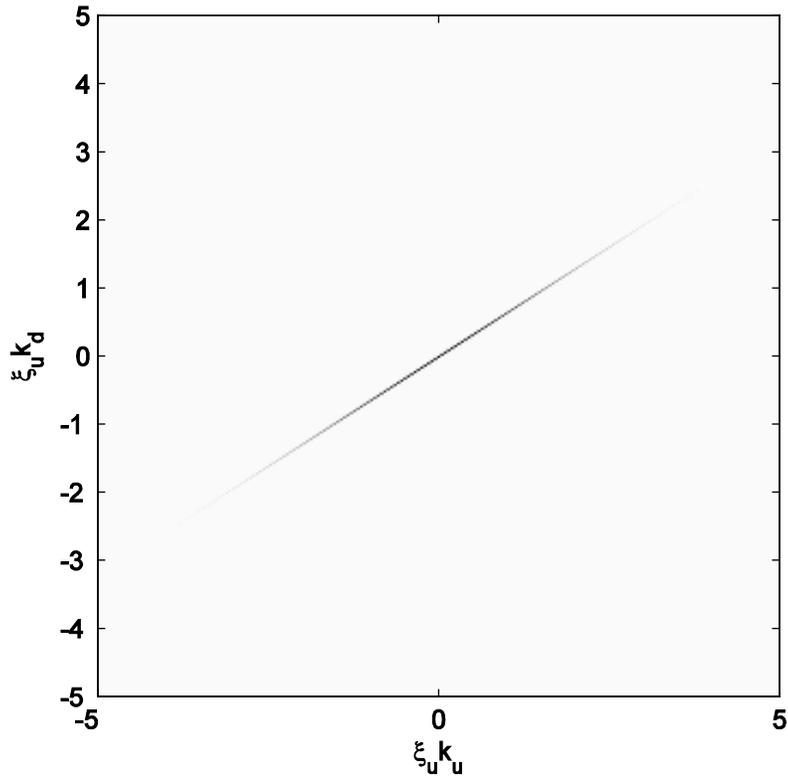

Fig. 2. The correlations between the Hawking and partner particles seen in the Fourier transform of the $(u,d)$ quadrant.

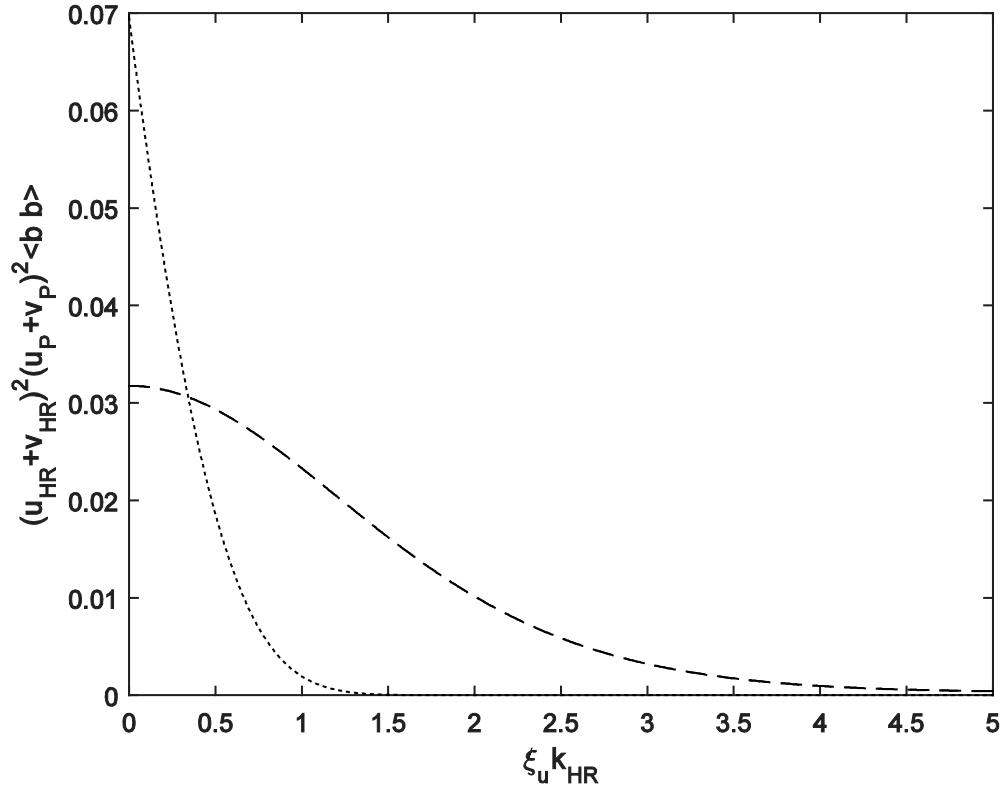

Fig. 3. The two terms in the entanglement parameter. The dashed curve indicates the correlations between the Hawking and partner particles. It is proportional to the Fourier transform along the dashed line of Fig. 1. The dotted curve shows the population of the Hawking particles (the Planck distribution). It is a linear function of the Fourier transform along the dotted line of Fig. 1 (the static structure factor). The dashed curve exceeding the dotted curve corresponds to entanglement.

By (9) and (10), the Fourier transforms of the $(u, u)$ and $(d, d)$ quadrants should be computed as well. This is equivalent to computing the 1D Fourier transforms along the dotted lines of Fig. 1. By (5) and (6), these Fourier transforms give the populations of the Hawking and partner particles, which should be same. The Hawking/partner population is indicated by a dotted curve in Fig. 3.

It is seen in Fig. 3 that the correlations between the Hawking and partner particles is weaker but broader than the Planck distribution. Thus, the Hawking/partner correlations exceed the Planck distribution for large $k$ only, so these $k$-values are entangled. This is expressed in Fig. 4 which shows the the Δ parameter, the difference between the two terms shown in Fig. 3. Positive values of $-\Delta$ correspond to entanglement. It is seen that the $k$-values in the high energy tail of the Planck distribution are entangled, whereas the low $k$ are not. This is consistent with the numerical results of Ref. 16.

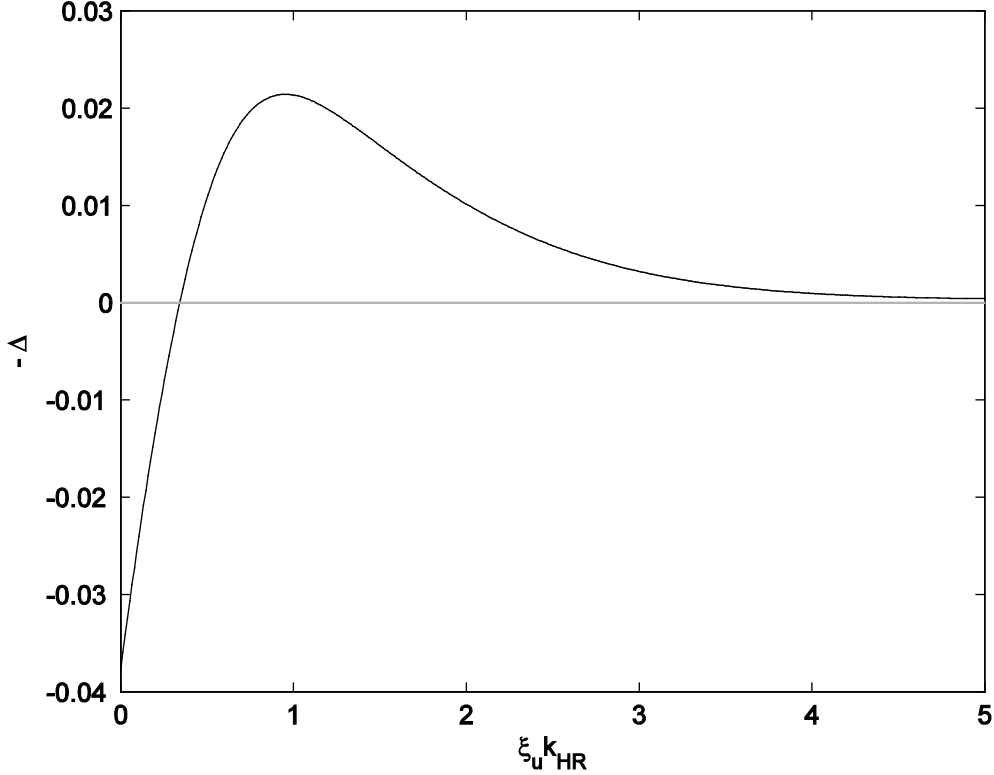

Fig. 4. The entanglement parameter. Positive values of $-\Delta$ correspond to entanglement.

For the sake of creating the figures above, we have utilized a specific type of acoustic black hole in a Bose-Einstein condensate. For the (u, d) and (d, u) quadrants of Fig. 1, we use the analytic result of Ref. 10 derived in the hydrodynamic limit, for a black hole with varying speed of sound but constant flow velocity. However, the general form of Fig. 1 is common to a variety of configurations [15], so the results here seem general. For the (u, u) and (d, d) quadrants of Fig. 1, we employ the well-known result that the density-density correlation function is the inverse Fourier transform of the static structure factor. Firstly, we assume that in the hydrodynamic limit, the populations are given by the Planck distribution $\langle \hat{b}^{u\,\dagger}_{k_{HR}} \hat{b}^{u}_{k_{HR}} \rangle = \langle \hat{b}^{d\,\dagger}_{k_P} \hat{b}^{d}_{k_P} \rangle = 1/[\exp(\hbar\omega/k_B T_H) - 1]$. We then insert this expression in (5) and (6) to obtain the static structure factors in the upstream and downstream regions $\langle \rho^{u}_{k_{HR}} \rho^{u}_{-k_{HR}} \rangle$ and $\langle \rho^{d}_{k_P} \rho^{d}_{-k_P} \rangle$. Finally we obtain the density-density correlation function by computing the inverse Fourier transform of (9) and (10).

In an experiment this process would be reversed, in that the density-density correlation function would be measured, the Fourier transform computed to obtain the static structure factor, and then the populations would be extracted from the static structure factor, as indicated by the dotted curve of Fig. 3. By graphing the quantities involved, we will see one of the major difficulties in measuring the entanglement. Fig. 5 shows the profile of the normally-ordered density-density correlation function along the dotted line of Fig. 1. It is seen that the population of Hawking particles barely changes the profile, since it is dominated by the quantum fluctuations of the condensate, represented by the unity term in (5) and (6). Fig. 6 shows the static structure factor, which has the same difficulty. The figures show the maximum possible contribution of the Hawking radiation, since we have chosen the maximum possible Hawking temperature (1/4 of the chemical potential) [15]. Thus, it is more difficult to measure the population of the Hawking particles than to measure $\langle \hat{b}^u_{k_{HR}} \hat{b}^d_{k_P} \rangle$, since the former is masked by background fluctuations.

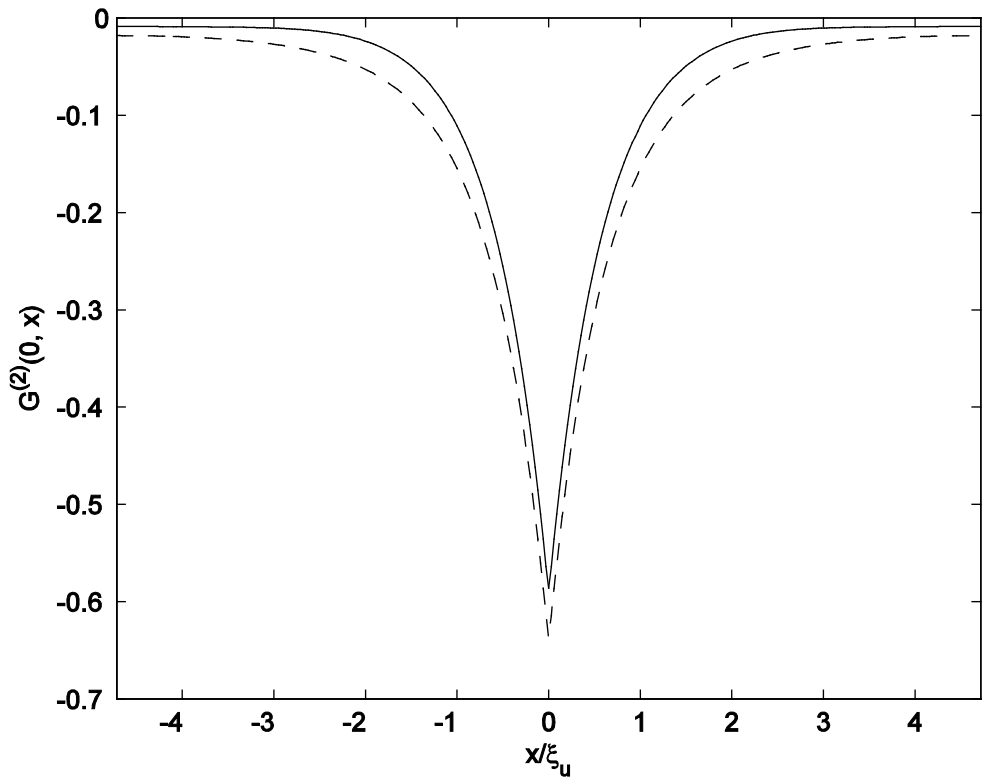

Fig. 5. The profile of the density-density correlation function. The solid curve includes the Hawking radiation, while the dashed curve does not. The dashed curve agrees with the curve in Ref. 15, obtained by other means.

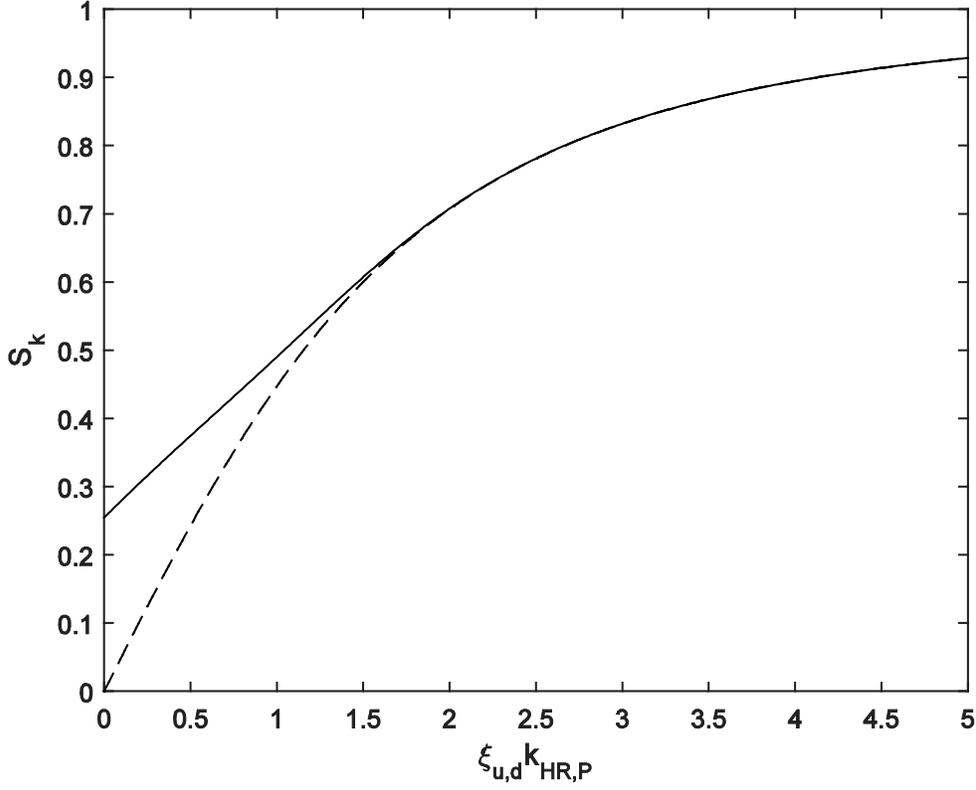

Fig. 6. The static structure factor. The solid curve includes the Hawking radiation, while the dashed curve does not.

We now turn to the Peres-Horodecki criterion, which is more general than (1). The criterion is based on the following quantity [16]

$$\mathcal{P}_- \equiv \left(\langle \hat{b}_{k_P}^{d\dagger}\hat{b}_{k_P}^d\rangle\langle \hat{b}_{k_{HR}}^{u\dagger}\hat{b}_{k_{HR}}^u\rangle - |\langle \hat{b}_{k_{HR}}^u\hat{b}_{k_P}^d\rangle|^2\right)\left[(1 + \langle \hat{b}_{k_P}^{d\dagger}\hat{b}_{k_P}^d\rangle)(1 + \langle \hat{b}_{k_{HR}}^{u\dagger}\hat{b}_{k_{HR}}^u\rangle) - |\langle \hat{b}_{k_{HR}}^u\hat{b}_{k_P}^d\rangle|^2\right] -$$

$$|\langle \hat{b}_{k_{HR}}^{u\dagger}\hat{b}_{k_P}^d\rangle|^2 \left(\langle \hat{b}_{k_{HR}}^{u\dagger}\hat{b}_{k_{HR}}^u\rangle + \langle \hat{b}_{k_P}^{d\dagger}\hat{b}_{k_P}^d\rangle + \langle \hat{b}_{k_{HR}}^{u\dagger}\hat{b}_{k_{HR}}^u\rangle\langle \hat{b}_{k_P}^{d\dagger}\hat{b}_{k_P}^d\rangle + 2|\langle \hat{b}_{k_{HR}}^u\hat{b}_{k_P}^d\rangle|^2\right) -$$

$$2\mathrm{Re}(\langle \hat{b}_{k_{HR}}^u\hat{b}_{k_P}^d\rangle^*\langle \hat{b}_{k_{HR}}^u\hat{b}_{k_{HR}}^u\rangle\langle \hat{b}_{k_{HR}}^{u\dagger}\hat{b}_{k_P}^d\rangle)(1 + 2\langle \hat{b}_{k_P}^{d\dagger}\hat{b}_{k_P}^d\rangle) - 2\mathrm{Re}\left(\langle \hat{b}_{k_{HR}}^u\hat{b}_{k_P}^d\rangle^{*2}\langle \hat{b}_{k_{HR}}^u\hat{b}_{k_{HR}}^u\rangle\langle \hat{b}_{k_P}^d\hat{b}_{k_P}^d\rangle\right) -$$

$$2\mathrm{Re}(\langle \hat{b}_{k_{HR}}^{u\dagger}\hat{b}_{k_P}^d\rangle\langle \hat{b}_{k_{HR}}^u\hat{b}_{k_P}^d\rangle\langle \hat{b}_{k_P}^d\hat{b}_{k_P}^d\rangle^*)(1 + 2\langle \hat{b}_{k_{HR}}^{u\dagger}\hat{b}_{k_{HR}}^u\rangle) + |\langle \hat{b}_{k_{HR}}^{u\dagger}\hat{b}_{k_P}^d\rangle^2 - \langle \hat{b}_{k_{HR}}^u\hat{b}_{k_{HR}}^u\rangle^*\langle \hat{b}_{k_P}^d\hat{b}_{k_P}^d\rangle|^2 -$$

$$|\langle \hat{b}_{k_P}^d\hat{b}_{k_P}^d\rangle|^2\langle \hat{b}_{k_P}^{d\dagger}\hat{b}_{k_P}^d\rangle(1 + \langle \hat{b}_{k_P}^{d\dagger}\hat{b}_{k_P}^d\rangle) - |\langle \hat{b}_{k_{HR}}^u\hat{b}_{k_{HR}}^u\rangle|^2\langle \hat{b}_{k_P}^{d\dagger}\hat{b}_{k_P}^d\rangle(1 + \langle \hat{b}_{k_P}^{d\dagger}\hat{b}_{k_P}^d\rangle). \quad (11)$$

If $\mathcal{P}_-$ is negative, then the Hawking radiation is entangled. In addition to (4), (5), and (6), the following 3 quantities are needed to evaluate $\mathcal{P}_-$:

$$\langle \rho^u_{k_{HR}} \rho^d_{-k_P} \rangle = \sqrt{N^u N^d}(u_{k_{HR}} + v_{k_{HR}})(u_{k_P} + v_{k_P})\langle \hat{b}^{u\,\dagger}_{k_{HR}} \hat{b}^d_{k_P} \rangle \qquad (12)$$

$$\langle \rho^u_{-k_{HR}} \rho^u_{-k_{HR}} \rangle = N^u(u_{k_{HR}} + v_{k_{HR}})^2 \langle \hat{b}^u_{k_{HR}} \hat{b}^u_{k_{HR}} \rangle \qquad (13)$$

$$\langle \rho^d_{-k_P} \rho^d_{-k_P} \rangle = N^d(u_{k_P} + v_{k_P})^2 \langle \hat{b}^d_{k_P} \hat{b}^d_{k_P} \rangle \qquad (14)$$

Similar to (8), (9), and (10), these three quantities are also given by

$$\langle \rho^u_{k_{HR}} \rho^d_{-k_P} \rangle = \int dx_1 dx_2 e^{-ik_{HR}x_1} e^{ik_P x_2} \langle n^u(x_1) n^d(x_2) \rangle \qquad (15)$$

$$\langle \rho^u_{-k_{HR}} \rho^u_{-k_{HR}} \rangle = \int dx_1 dx_2 e^{ik_{HR}x_1} e^{ik_{HR}x_2} \langle n^u(x_1) n^u(x_2) \rangle \qquad (16)$$

$$\langle \rho^d_{-k_P} \rho^d_{-k_P} \rangle = \int dx_1 dx_2 e^{ik_P x_1} e^{ik_P x_2} \langle n^d(x_1) n^d(x_2) \rangle \qquad (17)$$

Thus, $\mathcal{P}_-$ can be written in terms of the Fourier transforms of the various quadrants of the density-density correlation function, similar to $\Delta$. The Fourier transforms in (15)-(17) are computed parallel to the features in each quadrant of Fig. 1. Thus, they are zero, and $\langle \hat{b}^{u\,\dagger}_{k_{HR}} \hat{b}^d_{k_P} \rangle$, $\langle \hat{b}^u_{k_{HR}} \hat{b}^u_{k_{HR}} \rangle$, and $\langle \hat{b}^d_{k_P} \hat{b}^d_{k_P} \rangle$ are zero by (12)-(14). This results from the stationary, homogeneous form of the correlation function apparent in Fig. 1. The stationarity is seen in that the correlation feature in the (u, d) quadrant is independent of the distance from the origin (the distance from the horizon). The homogeneity is seen in that the features in the (u, u) and (d, d) quadrants are independent of the position along the diagonal. In this stationary homogeneous case, the Peres-Horodecki expression thus reduces to the first line of (11), which is equivalent to $\Delta$ in (1). This equivalence was also found in Ref. 16.

In conclusion, we see that the entanglement of Hawking radiation in a Bose-Einstein condensate can be measured from the density-density correlation function. This result is a significant experimental simplification. The method relies on the fact that the thermal populations of phonons are negligible, since the phonons traveling in the direction of the flow are neglected. Due to the background of quantum fluctuations, the measurement of the phonon populations is more difficult than the measurement of the correlations between the Hawking and partner particles. It is found that the Hawking/partner correlations are narrow in position space. They are therefore broad and weak in momentum space; even broader than the Planck distribution of Hawking and partner particles. Thus, the high-energy tail of the Planck distribution should be entangled but the low energies should not be entangled, in agreement with previous numerical results. The study of the entanglement of Hawking radiation in an analog system will hopefully shed light on the physics of real black holes.